\documentstyle[12pt]{article}

\begin{document}

\baselineskip=7mm

\begin{large}

\begin{center}

 {$J$-pairing Interactions of Fermions in a Single-$j$ Shell}

\vspace{0.1in}
    
        A. Arima

\vspace{0.2in}
                  
The House of Councilors, 2-1-1 Nagatacho,

   Chiyodaku, Tokyo 100-8962,  Japan

\vspace{0.315in}

\end{center}
\end{large}

\vspace{0.1in}

\begin{small}
In this talk I shall introduce our recent works on
general pairing interactions and pair truncation approximations
for fermions in a single-j shell, including the 
spin zero dominance, features of eigenvalues
of fermion systems in a single-j shell
interacting by a $J-$pairing interaction.

\vspace{0.4in}

\end{small}

Keywords : ~ $J$-pairing interaction, sum-rules, spin zero  dominance
              
\newpage

\section{Introduction}  

It is my great pleasure to talk to you here. I would like to thank
the organizers, especially Dr. Bruno Gruber.  I am extremely glad to
see many of my friends again today in this beautiful city Bregenz.

My talk consists of four subjects.

(1) Spin $0^+$ ground state dominance

(2) Pair approximations for fermions in a single-$j$ shell

(3) Regularities of states in the presence of $J_{\rm max}$-pairing

(4) Solutions for cases of $n=3$ and 4 with $H_J$
 
\section{$0^+$ ground state dominance}

A preponderance of $0^+$ ground states was discovered by
Johnson, Bertsch and Dean in 1998  \cite{Johnson1}
using the two-body random ensemble (TBRE), and was
related to a reminniscence of generalized seniority
by Johnson, Bertsch, Dean and Talmi in 1999 \cite{Johnson2}. 
These phenomena have been confirmed in different systems
\cite{Bijker1,Kusnezov}.

Let us take a simple systems consisting of four particles in a single-$j$
shell. The Hamiltonian that we use is as follows. 

\begin{eqnarray}
&&  H_J =
 \sum_{ J} G_J A^{J \dagger} \cdot  A^{J } \equiv \sum_J \sqrt{2J+1}
\left( A^{J \dagger} \times  A^{J }  \right)^{(0)} ~,
\nonumber \\
&&  A^{J \dagger} = \frac{1}{\sqrt{2}} \left[ a_{j}^{\dagger}
\times a_{j}^{\dagger}
     \right]^{(J)} ~,   
     {A}^J = - (-1)^M\frac{1}{\sqrt{2}} \left[ \tilde{a}_{j} \times
     \tilde{a}_{j} \right]^{(J)}.     
\end{eqnarray}
where $G_J$ is given by
\begin{eqnarray}
&& G_J = \langle j^2 J|V|j^2 J \rangle ~. \nonumber 
\end{eqnarray}                            
Here $V$ is a two-body interaction. 

We have used a two-body random ensemble to confirm this
interesting phenomenon and discovered an empirical method
to predict the probability of a ground state to have a
spin $I$ \cite{Zhao1}.
We keep only one $G_J$ to be $-1$ and all others 0:
\begin{eqnarray}
G_{J'} = - \delta_{JJ'} ~. \nonumber 
\end{eqnarray}
We then diagonalize the Hamiltonin to find the angular momenta
which give the lowest eigenvalues. They are shown in Table I.
We count how many different $G_J$'s give the largest eigenvalue to
an angular momentum $I$. The number is denoted as ${\cal N}_I$.
For example for $j=\frac{21}{2}$ and $n=4$,
${\cal N}_0$=5, ${\cal N}_2$=${\cal N}_4$=${\cal N}_8
={\cal N}_{20}$=${\cal N}_{28}$=${\cal N}_{36}$=1 and
all others are equal to 0. The total number of different
$G_J$'s is $N=\frac{2j+1}{2}$. Then the
$I$ g.s. probability is approximately predicted as
\begin{equation}
P^{\rm pred}(I) = {\cal N}_I/N.  \label{empirical}
\end{equation}

Fig. 1 shows a comparison between
$P^{\rm pred}(0)$ and $P^{\rm TBRE}(0)$ which is obtained by
diagonalization of a TBRE Hamiltonian for four fermions
in a  single-$j$ shells.
Fig. 2 shows a comparison of
between $P^{\rm pred}(I)$ and
$P^{\rm TBRE}(I)$ for examples of   various systems.

One can see that the agreements between the
 $P^{\rm pred}(I)$ and
$P^{\rm TBRE}(I)$ are very good. It is therefore important
to diagonalize $H$ with $G_{J'} = -\delta_{JJ'}$. For this purpose
we introduce the $J-$pair approximation for low-lying states.

\section{Pair Approximation for Fermions in a single-$j$ shell}

Our Hamiltonian is defined as
\begin{equation}
H_J = - A^{J \dag } \cdot A^J ~.
\end{equation}

We first point out that the low-lying eigenvalues of $H_J$
can be approximated by wavefunctions of pairs with spin $J$.
\begin{equation}
\Phi (I) = \frac{1}{\sqrt{N}} \left[
A^{J \dag } \times A^{J \dag } \times \cdots \times A^{J \dag }
\right]^{(I)} |0 \rangle ~,
\end{equation}
where $\frac{1}{ \sqrt{N}}$ is the normalization factor. It is
very easy to prove that the $J$-pair truncation describes the
low-lying states exactly in three body systems.

Fig. 3(a) shows the spin of the ground state of $j^4$ configuration
for $G_6=-1$. The ground states with spin 0 are obtained by
exact shell model calculations and by the $J$-pair approximation.
Fig. 3 (b) shows the similar thing for $G_{14}=-1$.

Fig. 4 shows energy levels obtained by shell model calculation and
by the $J$-pair approximation when $j=25/2$, $J=14$ and $n=4$. For
the low-lying states, the pair approximation is very good.
Giving the four low-lying states, two of them compete to be the ground
state. These energies are almost the same in both exact shell model
calculation and the pair approximation.  This
is why we failed to predict the ground state in this case.

For the $n$=5 and 6 cases that we have
examined, the low-lying states are reasonably well
approximated by the $J$-pair truncation.

So far $J$ is general, between 0 and $2j-1$. Now let us take a
very special value, $J_{\rm max} = 2j-1$.
For $H=H_{\rm max}= H_{2j-1}$, the $I= I_{\rm max} = 4j-6$
is the lowest, and $I=I_{rm max} -2 $ state is the second lowest.
These two states can be constructed by using pairs with angular momentum
either $J_{\rm max}$
or $J_{\rm max}-2$.

However,  pairs with angular momentum $J_{\rm max}-2$  do not
present a good approximation of the other $I$ states, while
those with angular momentum $J_{\rm max}$ do. For example, for $n=4$, 
$|J^{2}_{Jmax}, I=0 \rangle$ is exact but 
$|{(J_{max}-2)}^2, I=0 \rangle  $ is not exact, 
$|J^{2}_{\rm max}, I (\le j) \rangle$ is almost exact $(\sim -2)$ but  
$|{(J_{\rm max}-2)}^2, I (\le j) \rangle $ are not. 

\section{Regularities of states in the presence of $H_{J_{\rm max}}$ }

We first point out that
eigenvalues of low $I$ states ($n=3, 4, 5$) are approximately
integers. This can be proved  in terms of six-$j$ symbols
for $n=3$ \cite{Zhao2}.
For $n=4$, one can prove this 
  in terms of nine-$j$ symbols \cite{Zhao3}.

Another regularity may be examplified below by $j=21/2$ and $n=3$ and 4.
Among many states of $n=4$ with the same $I$, the lowest 
eigenvalue is expressed as ${\cal E}_I$ (obtained
by a shell model diagonalization).
The    ${\cal E}_I$ of 
four fermions in a single-$j$ ($j=21/2$) shell with 
$I$ between 18 to 25 are as follows. (When $I$ is smaller 
than 18 there is no eigenvalue lower than $-2$.) 
The eigenvalue of the $I^{(3)}_{\rm max} (=3j-3)$ state with three fermions 
in the same single-$j$ shell is $-\frac{59}{26}$=$-2.26923076923077$.
From Table II, one sees that
the  ${\cal E}_I$'s  of $n=4$ with 18$\le I \le 25$ are very
close to $E_{I^{(3)}_{\rm max} }$ and also very close to
that of an $I$ state constructed  by 
$\Psi_I =\left[ a_j^{\dag}\times \left[a_j^{\dag} \times
 a_j^{\dag} \times a_j^{\dag} \right]^{(I^{(3)}_{\rm max})} \right]^{(I)}$.

We have calculated overlaps between the above states of $n=4$
and the $\Psi_I$. They are almost 1 within a precision
of $10^{-5}$.
This phenomenon have been confirmed  for $n$ up to 6 ($j \ge 11/2$).

\section{Solutions for the case of $n=3$}

We take the following   basis  for three fermions 
\begin{eqnarray}
&& |j^3 [j J]  I, M \rangle =
     \frac{1}{\sqrt{ N^{(I)}_{j J; j J}}}
     \left(  a^{\dag}_j  \times
A^{J \dag} \right)^{(I)}_M |0 \rangle, \label{basis0} \nonumber
\end{eqnarray}
where $N^{(I)}_{j J; j J} $ is the diagonal matrtix element of the
normalization matrix
\begin{eqnarray}
&& N^{(I)}_{j J^{\prime}; j J} = \langle 0 |\left(  a_j
\times A^{J'} \right)^{(I)}_M  \left(  a_j^{\dag}  \times
A^{J\dag} \right)^{(I)}_M |0 \rangle.     \nonumber
\end{eqnarray}
In general this basis is over complete and the normalization matrix
may have zero eigenvalues for a given $I$. Here $J$ is not necessarily
equal to $J_{\rm max}$.

The $N^{(I)}_{j J^{\prime}; j J}$  and $
 \langle j^3 [j  K^{\prime}] I,M |
H_J  |j^3 [j K] I,M\rangle $ can be
evaluated analytically:    
\begin{eqnarray}
&& N^{(I)}_{j J^{\prime}; j J} 
=\delta_{J', J}  + 2 \hat{J}      \hat{J^{\prime}} 
     \left\{ \begin{array}{ccc}
     J    & j  & I \\
     J^{\prime}  & j & j . \end{array} \right\} , \nonumber \\
&&      \langle j^3 [j  K^{\prime}] I,M |
H_J  |j^3 [j K] I,M\rangle  
= -\frac {1}{{\sqrt{N^{(I)}_{j K^{\prime}; j K^{\prime}}
N^{(I)}_{j K; j K} } }}
  N^{(I)}_{j  K^{\prime}; j J}
  N^{(I)}_{j J; j  K}  .  \nonumber
\end{eqnarray}
where $\hat{L}$ is a short hand notation of $\sqrt{2L+1}$.

For a fixed $J$ and   any $I$, we construct one state
$|j^3 J: I \rangle= |j^3 [jJ ]I \rangle$ and all other 
states $|j^3 K: I \rangle$, which are orthogonal to 
$|j^3 J: I \rangle$, as follows:
\begin{eqnarray}
&& |j^3 K: I \rangle= |j^3 [j K] I \rangle
- \frac
{N^{(I)}_{j K; j J}}{\sqrt{N^{(I)}_{j J; j J} N^{(I)}_{j K; j K}}}|j^3
[jJ]I \rangle,\ (K \neq J), \nonumber \\
&&      |j^3 J: I \rangle= |j^3 [jJ ]I \rangle. \label{basis} \nonumber 
\end{eqnarray}

One easily confirms that all matrix elements of 
the Hamiltonian,  
$ \langle j^3 K':I | H_J  | j^3 K: I \rangle$,  are 
zero,  except for $K' = K = J$: 
\begin{eqnarray}
\langle  j^3 [jJ ]I | H_J|j^3 [jJ ]I \rangle
= - N^{(I)}_{j J; j J} = -1 -2 (2J+1) 
     \left\{ \begin{array}{ccc}
     J    & j  & I \\
     J    & j & j . \end{array} \right\} ~ .   \nonumber 
\end{eqnarray}

Thus,   all the eigenvalues of $H_J$ for $n = 3$ with any angular momentum $I$
 are zero except for the 
state with one pair of  spin  $J$, which has the  eigenvalue
$E_I^{J(j)}=- N^{(I)}_{jJ; jJ}$. 

As by-products, we obtain a number of sum rules for six-$j$ symbols.                          
The procedure to derive these sum rules is straightforward. As is 
well known, 
the summation of all eigenvalues with a fixed $I$ is equal to
$\frac{n(n-1)}{2}$ times the number of $I$ states, where $n$ is
the particle number.  
For $n=3$, the number of states can be expressed in a compact
formula \cite{nucl-th/0304038}.

We use $E_I^{J(j)}$ to denote the non-zero eigenvalue of 
  $H= H_J$ for any $I$ , we have that 
\begin{eqnarray}
&& \sum_J E_I^{J(j)} = \sum_J \left[
-1 - 2 (2J+1) 
   \left\{ \begin{array}{ccc}
    j    & I  & J \\
    j    & j  & J  \end{array} \right\} \right] \nonumber \\
&& ~~~~ ~~~~~ ~~~   =
- \frac{n(n-1)}{2} D(j^3, I) ~. \nonumber
\end{eqnarray}

For  $I \le j$ ($j$ is a half integer), 
\begin{eqnarray}
&& \sum_{J={\rm even}} 2(2J+1)
    \left\{ \begin{array}{ccc}
    j    & I  & J \\
    j    & j  & J  \end{array} \right\}
=3 \left[ \frac{2I+3}{6} \right] - I - \frac{1}{2} ~, \nonumber
\label{new1}
\end{eqnarray}
where $\left [ ~~ \right]$ means to take the largest  integer not exceeding
the value inside. 

Our new sum rules of six-$j$ symbols will be given in \cite{Zhao3}
in details. 

\section{Summary}

In this talk, I have discussed four interesting
aspects  concerning 
general pairing interactions and pair
truncation approximations for fermions in a single-$j$ shell.
I first discuss an empirical rule to predict the spin $I$
ground state probability. 
I then show that pairs with spin $J$ are reasonable building blocks
for the low-lying states of a Hamiltonian with an attractive
$J$-pairing interaction only. I also present two interesting regularities
of eigenvalues of Hamiltonian with $J_{\rm max}$-pairing interaction:
for low $I$ states of $n$ up to 5 we found that the eigenvalues
are asymptotic integers;  some of $n=4$ states may be tracted back
to $n=3$. Finally I prove   for the case of $n=3$ the eigenvalues
are written in terms of six-$j$ symbols. This result presents 
new sum rules of six-$j$ symbols. 

Acknowledgements: 
I would like to thank Drs. Y. M. Zhao, J. N. Ginocchio, and
N. Yoshinaga for their collaborations in this work.

\newpage

Figure Captions:

\vspace{0.3in}

Figure ~ 1 ~~
Comparison between $P^{\rm pred}(0)$ and $P^{\rm TBRE}(0)$
of four fermions in a single-$j$ shell.
The solid squares are obtained by 1000 runs of a
TBRE Hamiltonian and the open squares are predicted
by Eq. (\ref{empirical}).

\vspace{0.3in}

Figure ~ 2 ~~ Comparison between $P^{\rm pred}(I)$ and $P^{\rm TBRE}(I)$
for more complicated systems.
The solid squares are obtained by 1000 runs of a
TBRE Hamiltonian and the open squares are predicted
by Eq. (\ref{empirical}).

\vspace{0.3in}

Figure ~ 3 ~~ Ground state spin $I$ for four fermions in a single-$j$ shell for
 $J=6$ in (a) and 14 in (b) as a function of $j$. The solid
   squares are obtained by diagonalized the $H_J$ in the
 full shell model space and open squares are obtained by truncating the space
 with two pairs with spin $J$ only.

\vspace{0.3in}

Figure ~ 4 ~~ 
A comparison of low-lying spectra
with  two pairs with spin $J=14$ (the column on the left hand side)   and by a diagonalization
of the full space (the column  in the middle  and the column on the right hand side) for
the case of four nucleons in a single-$j$ ($j=25/2$) shell.
The middle column plots the shell model states which are well reproduced
by the two $J=14$ pairs, and the right column plots
the shell model states which are not well reproduced by two $J=14$ pairs.
All the levels below $0^+_1$ in the full shell model
space  are included.
One sees that the low-lying states
with $I=2_1^+$, $6_1^+$, $12_1^+$, and $10_1^+$
are well reproduced.   
 
\newpage

Table ~ I ~~ The angular momenta which give the lowest eigenvalues
when $G_J=-1$  and all other parameters are  0 for 4 fermions in
single-$j$ shells.  

\vspace{0.3in}

\begin{tabular}{ccccccccccccccccc} \hline  
$2j$ &  $G_0$ &  $G_2$ &  $G_4$ &  $G_6$ &  $G_8$ &  $G_{10}$ &  $G_{12}$
&  $G_{14}$ &  $G_{16}$ &  $G_{18}$ &  $G_{20}$
  \\  \hline
7  & 0 &4 &2 &8 &   &   & & & &  &  \\
9  & 0 &4 &0 &0 &12 &   & & & &  &  \\
11 & 0 &4 &0 &4 &8  &16 & & & &  &  \\
13 & 0 &4 &0 &2 &2  &12 &20 & & &  &  \\
15 & 0 &4 &0 &2 &0  &0  &16 &24 & &  &  \\
17 & 0 &4 &6 &0 &4  &2  &0  &20 &28 &  &  \\
19 & 0 &4 &8 &0 &2  &8  &2  &16 &24 &32 &  \\
21 & 0 &4 &8 &0 &2  &0  &0  &0  &20 &28 &36  \\   \hline   
\end{tabular}
 
\vspace{0.3in}

Table ~ II ~~ 
A comparison between eigen-energies
obtained by diagonalizing $H_{J_{\rm max}}$ in the full shell model space 
(the column ``(SM)")
and matrix elements
$\langle \Psi_I | H | \Psi_I \rangle$ (column ``$F_I$").

\vspace{0.3in}

\begin{tabular}{c|cc} \hline 
 $I$ & ${\cal E}_I$ (SM) ~  & $F_I$ (coupled)  \\ \hline
18 & -2.26923076925915  & -2.26923076923498     \\
19 & -2.26923076930701  & -2.26923076930702    \\
20 & -2.26923078555239  & -2.26923077167687    \\
21 & -2.26923078386646  & -2.26923078385375    \\
22 & -2.26923245245008  & -2.26923102362432    \\
23 & -2.26923165420128  & -2.26923165276669    \\
24 & -2.26930608933736  & -2.26924197057701    \\
25 & -2.26925701778767  & -2.26925692933680    \\
 \hline 
\end{tabular}

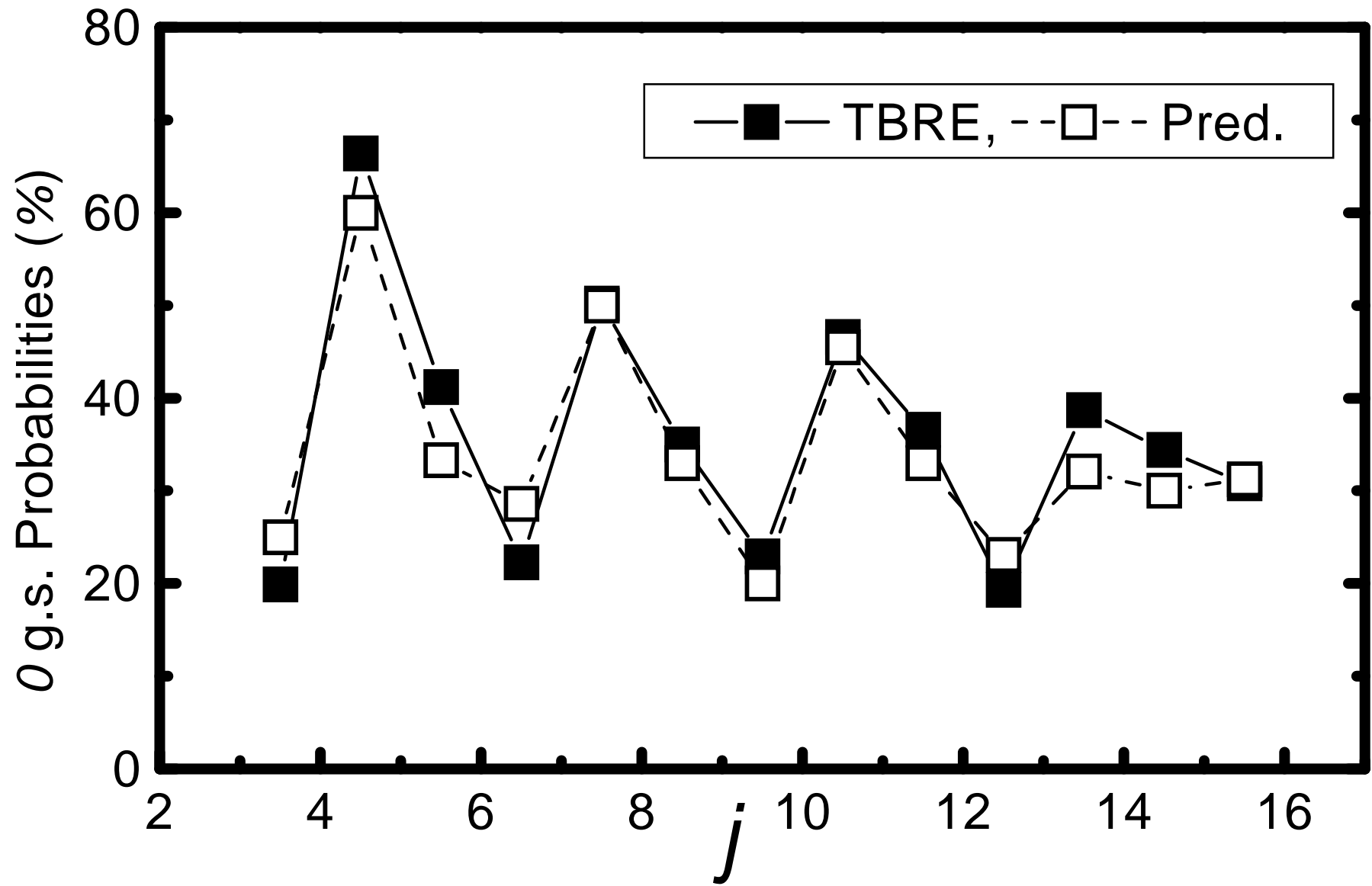
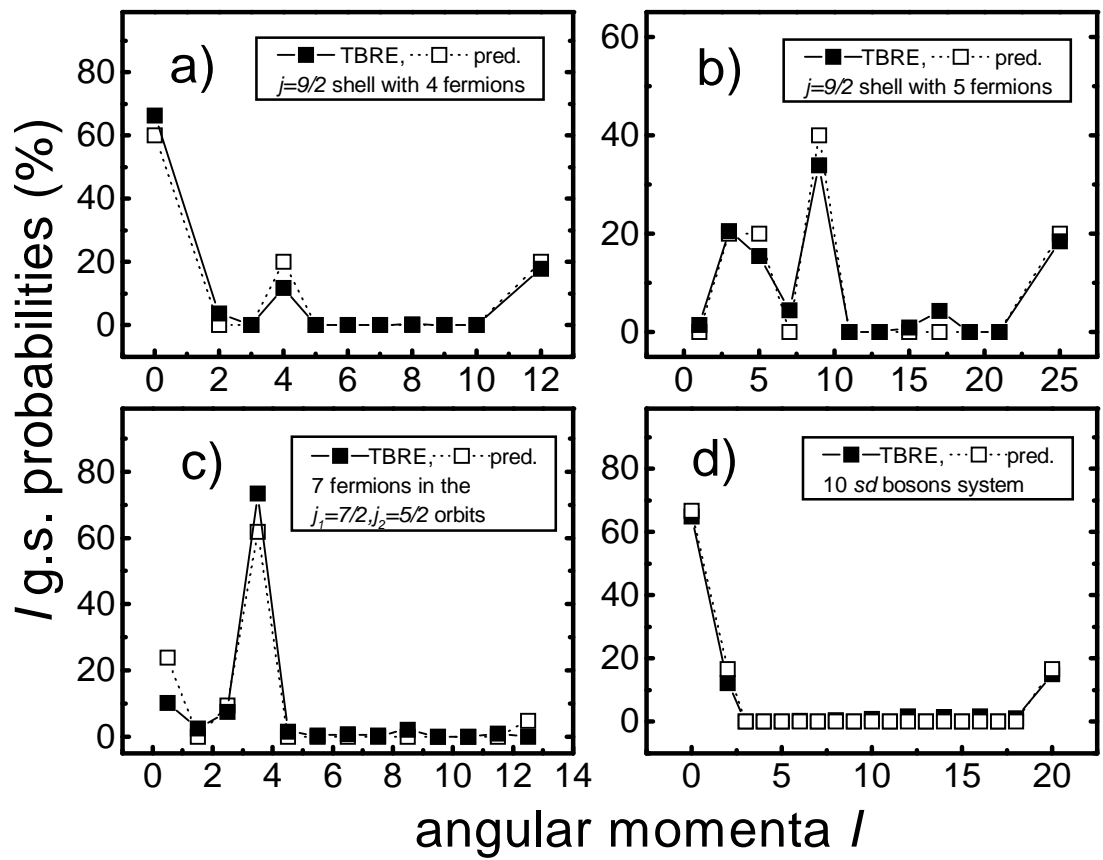
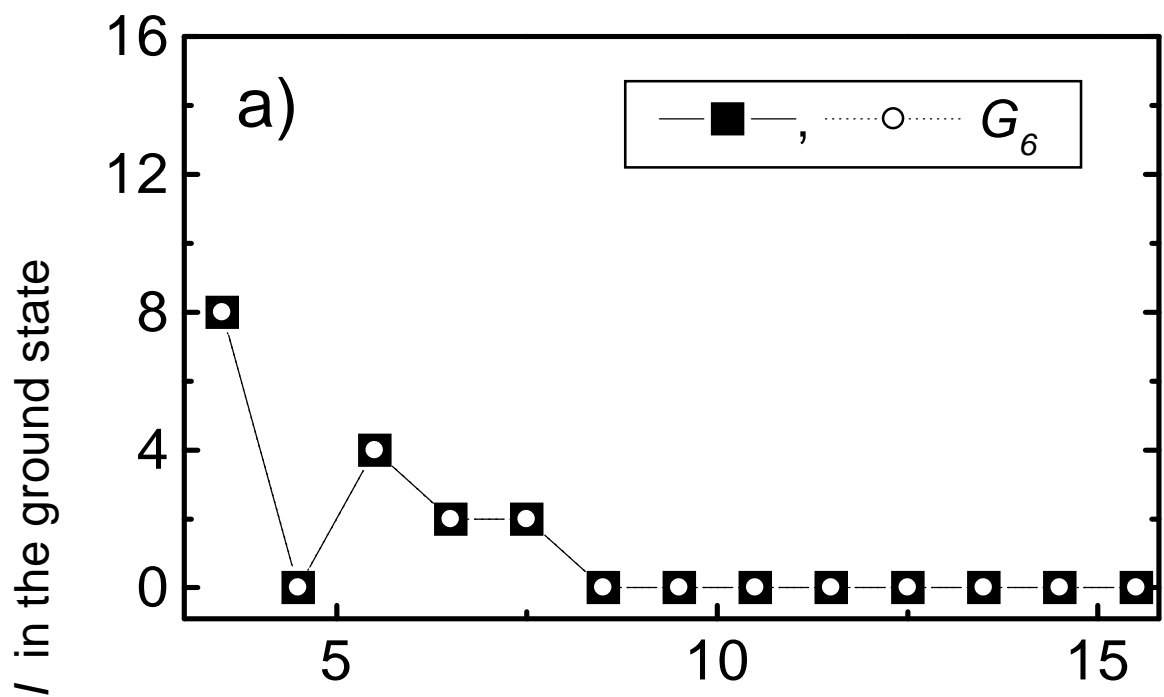
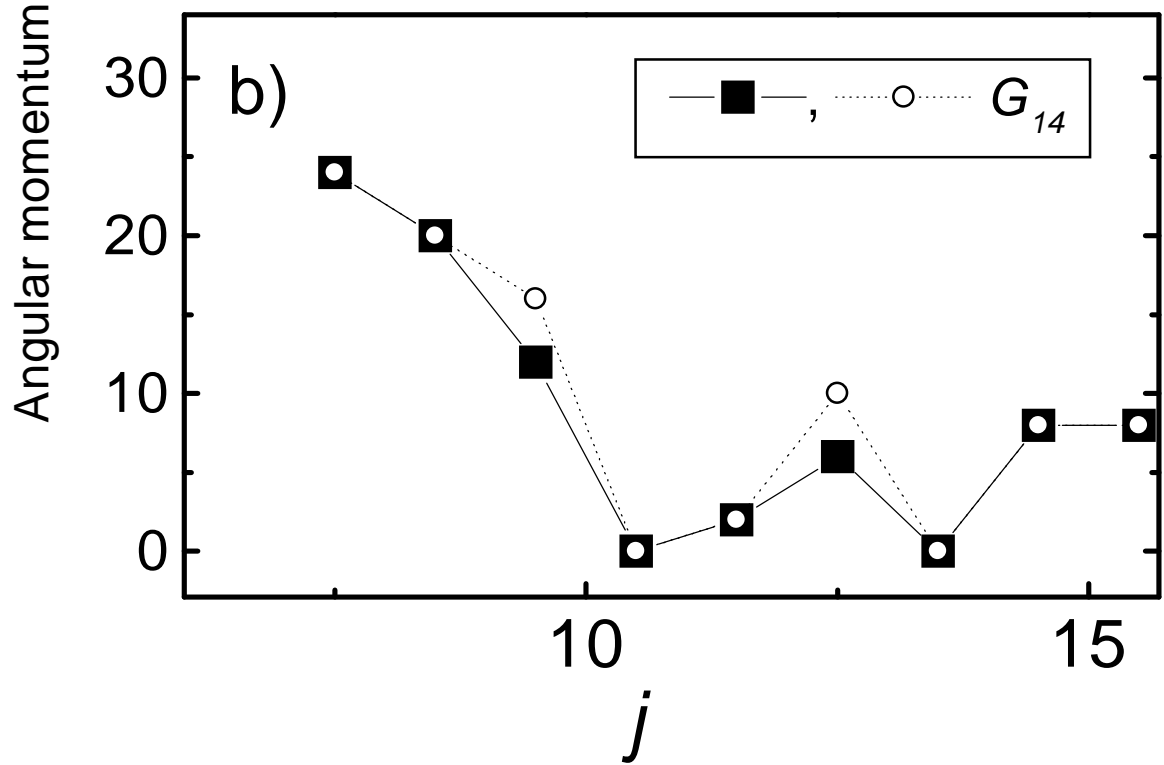
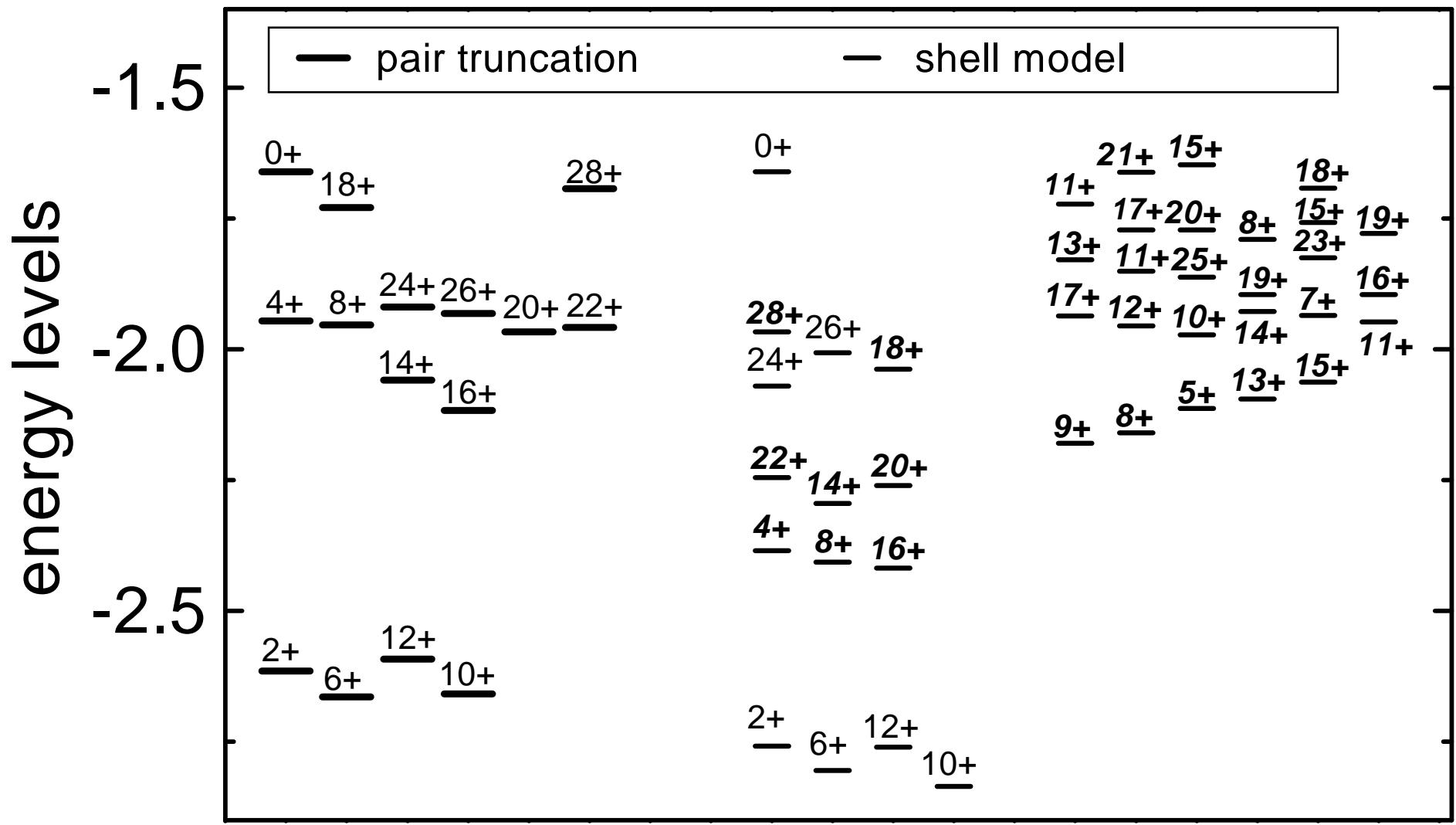
\end{document}